\documentclass{elsart}
\usepackage{graphics}
\begin{document}
\begin{frontmatter}
\title{On the effective light-cone QCD-Hamiltonian:\\
       Application to the pion and other mesons}
\author{Hans-Christian Pauli} 
\address{Max-Planck-Institut f\"ur Kernphysik, 
         D-69029 Heidelberg, Germany}
\begin{abstract}
  The effective interaction between a quark and an anti-quark
  as obtained previously with by 
  the method of iterated resolvents
  is replaced by the $\uparrow\downarrow$-model 
  and applied to flavor off-diagonal mesons including the $\pi^+$.
  The only free parameters are the canonical ones, the coupling
  constant and the masses of the quarks. 
\end{abstract}
\hyphenation{ }
\end{frontmatter}

The light-cone approach \cite{Dir49} to the bound-state problem 
in gauge theory \cite{BroPauPin98}
aims at solving  
$H_{LC}\vert\Psi\rangle = M^2\vert\Psi\rangle$.
If one disregards possible zero modes \cite{KalPau93-96}
and works in the light-cone gauge, 
the (light-cone) Hamiltonian $H_{LC}$ 
is given as a Fock-space operator in \cite{BroPauPin98},
or in the compendium \cite{Com00}. 
Its eigenvalues are the invariant mass-squares  $M^2$
of physical particles associated with the eigenstates
$\vert\Psi\rangle$.
In general, they are superpositions of all possible
Fock states with its many-particle configurations.
For a meson, for example, holds 
\begin{center}
\(\displaystyle  
\begin{array} {rcll} 
      {\vert\Psi_{\rm meson}\rangle} &=& {\sum\limits_{i}\ }
      {\Psi_{q\bar q}(x_i,\vec k_{\!\perp_i},\lambda_i)}
      &{\vert q\bar q\rangle} 
\\&+& {\sum\limits_{i}\ }
      {\Psi_{g g}(x_i,\vec k_{\!\perp_i},\lambda_i)}
      &{\vert g g\rangle }
\\&+& {\sum\limits_{i}\ }
      {\Psi_{q\bar q g}(x_i,\vec k_{\!\perp_i},\lambda_i)}
      &{\vert q\bar q g\rangle }
\\&+&  {\sum\limits_{i}\ }
      {\Psi_{q\bar q q\bar q }(x_i,\vec k_{\!\perp_i},\lambda_i)}
      &{\vert q\bar q q\bar q \rangle}
\\&+& {\dots}.
\end{array}\)
\end{center}
If all wave functions like $\Psi_{q\bar q}$ or $\Psi_{g g}$
are available, one can analyze hadronic structure 
in terms of quarks and gluons \cite{BroPauPin98}.~--
Hiller \cite{Hil00}, or Tang {\it et al.} \cite{TanBroPau91}, 
attacks this problem by DLCQ.
Alternatively, one addresses to reduce the many-body problem
behind a field theory to an effective one-body problem.
The derivation of the effective interaction becomes then the key issue.
By definition, an effective Hamiltonian acts only
in the lowest sector of the theory  
(here: in the Fock space of one quark and one anti-quark) 
and has the same eigenvalue spectrum
as the full problem.
I have derived such an effective interaction 
by the method of iterated resolvents \cite{Pau98}, that is
by systematically expressing the higher Fock-space
wave functions as functionals of the lower ones.  
In doing so the Fock-space is not truncated
and all Lagrangian symmetries are preserved.
The projections of the eigenstates onto the 
higher Fock spaces can be retrieved
systematically from the $q\bar q$-projection, 
with explicit formulas given in \cite{Pau99b}.
I have derived \cite{Pau00} the effective interaction also with 
the method of Hamiltonian flow equations \cite{Weg00}.

\section{The eigenvalue equation in the $q\bar q$-space}
\label{sec:1}

For flavor off-diagonal mesons
(to mesons with a different flavor for quark and anti-quark),
the effective one-body equation \cite{Pau98} is shockingly simple:
\begin{eqnarray} 
    &&M^2\langle x,\vec k_{\!\perp}; \lambda_{1},
    \lambda_{2}  \vert \psi\rangle  
    = \left[ 
    \frac{\overline m^2_{1} + \vec k_{\!\perp}^{\,2}}{x} +
    \frac{\overline m^2_{2} + \vec k_{\!\perp}^{\,2}}{1-x}  
    \right]\langle x,\vec k_{\!\perp}; \lambda_{1},
    \lambda_{2}  \vert \psi\rangle   
\label{eq:1}\\ 
    && - \frac{1}{3\pi^2}
    \sum _{ \lambda_q^\prime,\lambda_{2}^\prime} \!\int 
    \frac{ dx^\prime d^2 \vec k_{\!\perp}^\prime
    \,R(x',\vec k'_{\!\perp};\Lambda) }
    {\sqrt{ x(1-x) x^\prime(1-x^\prime)}}
    \frac{\overline\alpha}{\,Q^2} 
    \langle
    \lambda_{1},\lambda_{2}\vert S\vert 
    \lambda_{1}^\prime,\lambda_{2}^\prime\rangle
    \,\langle x^\prime,\vec k_{\!\perp}^\prime; 
    \lambda_{1}^\prime,\lambda_{2}^\prime  
    \vert \psi\rangle.  
\nonumber\end{eqnarray} 
Here, $M ^2$ is the eigenvalue of the invariant-mass squared. 
The associated eigenfunction $\psi\equiv\Psi_{q\bar q}$ 
is the probability amplitude 
$\langle x,\vec k_{\!\perp};\lambda_{1},\lambda_{2}\vert\psi\rangle$ 
for finding a quark with momentum fraction $x$, 
transversal momentum $\vec k_{\!\perp}$ 
and helicity $\lambda_{1}$,
and correspondingly the anti-quark with
$1-x$, $-\vec k_{\!\perp}$ and $\lambda_{2}$.
The $\overline m _1$  and $\overline m _2$ 
are (effective) quark masses  and 
$\overline\alpha$ is the (effective) coupling constant. 
The mean Feynman-momentum transfer of the quarks is
denoted by $Q^2$, 
\begin{equation}
   Q ^2 (x,\vec k_{\!\perp};x',\vec k_{\!\perp}') 
   = -\frac{1}{2}\left[(k_{1}-k_{1}')^2 + (k_{2}-k_{2}')^2\right]
,\end{equation}
the spinor factor $S=S(x,\vec k_{\!\perp};x',\vec k_{\!\perp}')$ by 
\begin{equation}
   \langle\lambda_{1},\lambda_{2}\vert S\vert
   \lambda_{1}^\prime,\lambda_{2}^\prime\rangle =
   \left[ \overline u (k_{1},\lambda_{1})\gamma^\mu
   u(k_{1}^\prime,\lambda_{1}^\prime)\right] \, 
   \left[ \overline v(k_{2}^\prime,\lambda_{2}^\prime) \gamma_\mu 
    v(k_{2},\lambda_{2})\right] 
.\label{eq:3}\end{equation}
The regulator function $R(x',\vec k'_{\!\perp};\Lambda)$ 
restricts the range of integration as function of some mass scale 
$\Lambda$.
I happen to choose here a soft cut-off (see below),
in contrast to the previous sharp cut-off \cite{KPW92,TriPau00}.  
Note that Eq.(\ref{eq:1}) is a fully relativistic equation.

The full QCD-Hamiltonian in 3+1 dimensions must be regulated
from the outset.
There are arbitrarily many possible ways to do that,
see {\it f.e.} Hiller \cite{Hil00}.
One of the few practical ways is vertex 
regularization \cite{BroPauPin98,Pau98},
where every Hamiltonian matrix element,
particularly those of the vertex interaction
(the Dirac interaction proper), 
is multiplied
with a convergence-enforcing momentum-dependent function,
which is kind of a form factor \cite{BroPauPin98}.
The precise form of this function is unimportant here,
provided it prevents that the scattered particles 
go too much off-shell
as function of a cut-off scale ($\Lambda$).
In the limit $\Lambda\rightarrow\infty$
the full unregulated theory is restored.
The effective quark masses $\overline m _1$ and
$\overline m _2$ and  
the effective coupling constant $\overline\alpha$ 
depend, in general, on $\Lambda$.
Explicit expressions for a sharp cut-off are available \cite{Pau98}.
In the spirit of renormalization theory they are renormalization
constants, subject to be determined by experiment,
and hence-forward will be denoted by $m _1$,
$m _2$, and $\alpha$, respectively.
In next-to-lowest order of approximation the coupling 
constant becomes a function of the momentum transfer, 
$\overline \alpha\longrightarrow\overline \alpha(Q;\Lambda)$, 
with the explicit expression given in \cite{Pau98}.

Why is Eq.(\ref{eq:1}) so shocking? 

In the first place, I am baffled by the fact that the effective 
interaction for QCD and QED differs only by the color factor
$(n_c^2-1)/(2n_c)=4/3$.
This is in raging conflict with all what I have learned
over the past 20 years on vacuum structure, condensates,
soft pion theorems, confinement, and the like. 
In fact, I knew the simple color factor already 
at the first light-cone meeting in 1991.
I have hidden it quite well in Ref.\cite{KPW92},
where I had obtained Eq.(\ref{eq:1}) by a light-cone adapted 
Tamm-Dancoff approach, which is an educated guess at the most.
Like everybody else, I could not believe that 
Eq.(\ref{eq:1}) is applicable to the pion, for example,
although it really should, since no assumption was made 
on the size of the quark masses. In consequence, 
I have spent in vain many painful years to accumulate 
counter evidence by a more rigorous formalism,
first by investigating 
the zero modes \cite{KalPau93-96},
then by developing the method of 
iterated resolvents \cite{Pau98}.
For some time around the 1998 meeting in St. Petersburg,
I believed that the salvation was in the functional 
behaviour of the running coupling constant 
$\overline \alpha(Q;\Lambda)$.
But without screwing the parameters beyond all reason,
I could not get the pion quantitatively (unpublished),
the pion that mystery particle of QCD.
Ultimately I am thrown back to Eq.(\ref{eq:1}),
the point of departure. 

But nowadays I am in a better shape. 
I have obtained Eq.(\ref{eq:1}) by two 
different but well founded methods:
by the iterated resolvents \cite{Pau98}
and by the Hamiltonian flow equations \cite{Weg00}.

It might be to early for solving Eq.(\ref{eq:1}) 
numerically in full glory like in Ref.\cite{TriPau00}.
Rather should I try to dismantle the equation of all irrelevant
details (or what I think they are)
and work as much as possible analytically.
With other words, I should develop a simple model.

\section{The $\uparrow\downarrow$-model and its application to the pion}
\label{sec:2}

In light-cone parametrization, the quarks are at relative rest
when $\vec k _{\!\perp}= 0$ and 
$ x = \overline x \equiv m_1/(m_1+m_2)$.
For very small deviations from these equilibrium values 
the spinor matrix is proportional to the unit matrix, with 
\begin{equation}
   \langle\lambda_1,\lambda_2\vert S\vert\lambda_1'\lambda_2'\rangle
   \sim 4 m_1 m_2 
   \ \delta_{\lambda_1,\lambda_1'}
  \ \delta_{\lambda_2,\lambda_2'}
,\end{equation}
see Compendium.   
For very large deviations, particularly for
$\vec k_{\!\perp}^{\prime\,2} \gg \vec k_{\!\perp} ^{\,2}$,
holds  
\begin{equation}
   Q ^2 \simeq\vec k_{\!\perp}^{\prime\,2} 
,\hskip2em\mbox{ and } \hskip2em
   \langle\uparrow\downarrow\vert S\vert\uparrow\downarrow\rangle 
   \simeq 2\vec k_{\!\perp}^{\prime\,2}
.\end{equation}
Both extremes are combined in the ``$\uparrow\downarrow$-model'':
\begin{equation}
   \frac{\langle\uparrow\downarrow\vert S\vert\uparrow\downarrow\rangle}
        {Q^2} 
   =
   \frac{\langle\downarrow\uparrow\vert S\vert\downarrow\uparrow\rangle}
        {Q^2} 
   \simeq 
   \frac{4 m_1 m_2}{Q ^2} + 2   
.\end{equation}
It interpolates between two extremes:
For small momentum transfer, 
the `2' is unimportant and the dominantly Coulomb aspects
of the first term prevail.
For large momentum transfers the Coulomb aspects are
unimportant and the hyperfine interaction is dominant.
But the model over-emphasizes many aspects: 
It neglects the momentum dependence of the Dirac spinors
and thus the spin-orbit interaction; it also neglects the
momentum dependence of the spin-spin interaction,
keeping only the `2' as its residue.
This `2' creates havoc: Its Fourier transform is a 
Dirac-delta function 
with all its consequences in a bound-state equation.

Here is an interesting point: We are all familiar
with the field theoretic divergences residing
in the effective masses 
and the effective coupling constant.
We are not used to ``divergences'' residing in a 
\underline{finite} number `2'.
They must also be regulated and renormalized.

I regulate thus the kernel by
\begin{equation}
   R(\mu)\ \frac{S} {Q ^2} \equiv 
   \frac{\mu^2}{\mu^2+Q^2}
   \left(\frac{4 m_1 m_2}{Q ^2} + 2\right) 
   \simeq
   \frac{4 m_1 m_2}{Q ^2} + 
   \frac{2\mu^2}{\mu^2+Q^2} 
.\label{eq:4}\end{equation}
I denote the soft cut-off by $\mu$
(in contrast to a sharp cut-off $\Lambda$).
The Coulomb interaction in first term 
needs no regularization.
In consequence I replace Eq.(\ref{eq:1}) by
\begin{eqnarray} 
    &M^2&
    \psi(x,\vec k_{\!\perp}) = \left[ 
    \frac{ m^2_{1} + \vec k_{\!\perp}^{\,2}}{x} +
    \frac{ m^2_{2} + \vec k_{\!\perp}^{\,2}}{1-x}  
    \right]\psi(x,\vec k_{\!\perp}) 
\nonumber\\ 
    &-&  \frac{\alpha}{3\pi^2} \!\int 
    \frac{ dx' d^2 \vec k_{\!\perp}'}
    {\sqrt{ x(1-x) x'(1-x')}}
    \left(\frac{4 m_1 m_2}{Q ^2} + 
    \frac{2\mu^2}{\mu^2+Q^2}\right)
    \psi(x',\vec k_{\!\perp}')
,\label{eq:20m}\end{eqnarray} 
where 
$\psi(x,\vec k_{\!\perp})\equiv
 \langle x,\vec k_{\!\perp}; \uparrow,\downarrow
 \vert \psi\rangle $.
On top of the canonical parameters $m$ and $\alpha$,
the eigenvalues depend now on a regularization scale $\mu$.

Since $\mu$ is an unphysical parameter, 
I must remove its impact.
I do that in the spirit we had treated the pairing
interaction in nuclei \cite{BraDamJen72}. 
It also has a Dirac-delta function.
As emphasized repeatedly by the late and un-forgotten 
Strutinky \cite{BraDamJen72},
one must adjust the (pairing) coupling constant
and the (pairing) cut-off such,
that the binding energy and the pairing gap 
would not depend on either of them.
I proceed therefore by looking for a cut-off dependent 
coupling function $\alpha_{\mu}$ such,
that the {\em calculated mass of the pion}  
agrees with the {\em empirical} value.
The so obtained function $\alpha_{\mu}$ is considered universal.
The {\em actual value} of $\mu$ must be fixed by a second requirement.
I emphasize that this procedure is not a renormalization 
group treatment in the modern sense; 
but it is a first step in the right direction.

\begin{figure} [t]  
\begin{minipage}[t]{68mm}
  \hspace{-2em}
  \resizebox{1.19\textwidth}{!}{\includegraphics{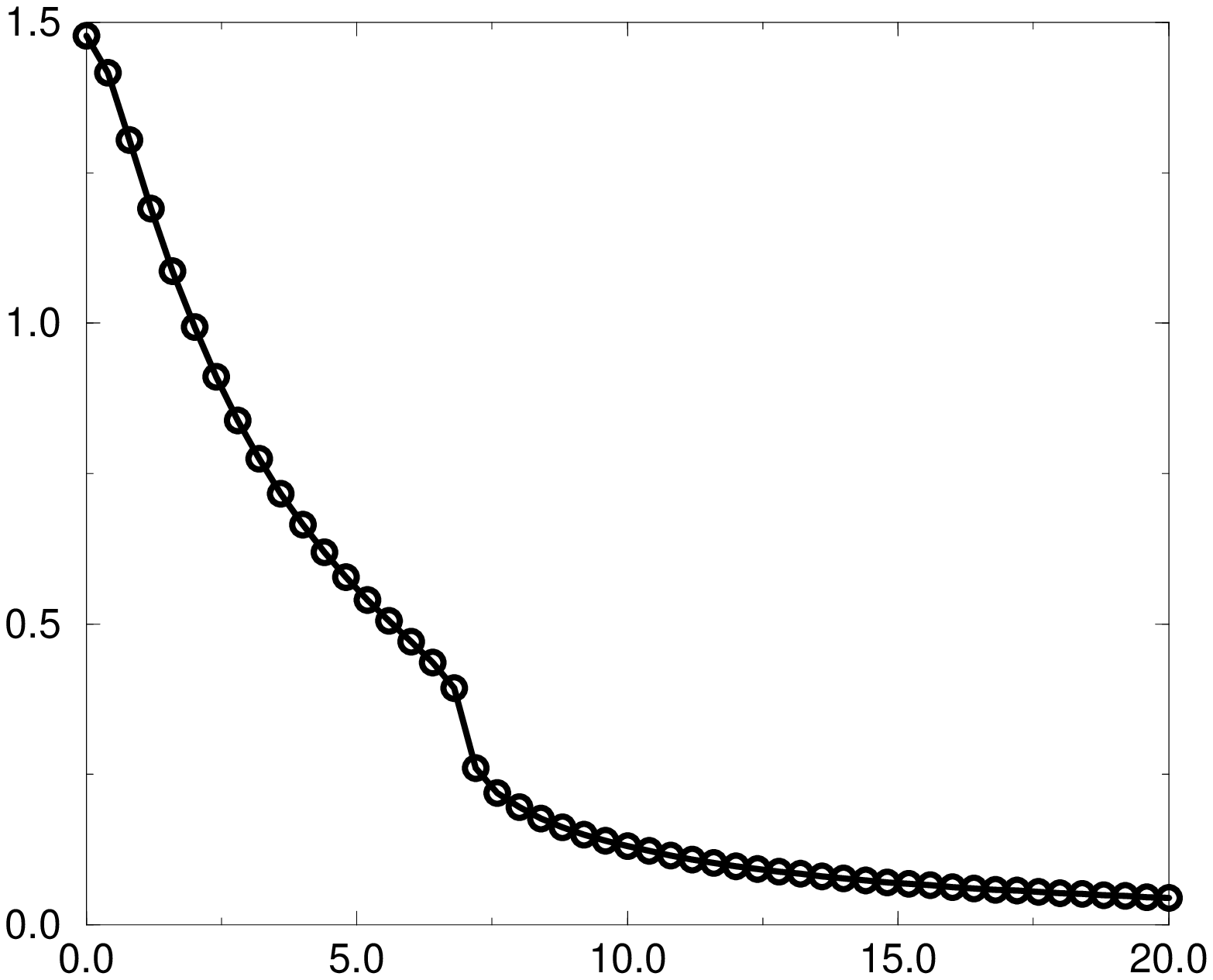}}
\caption{\label{fig:5.1}  
   Coupling constant $\alpha$ versus the regularization scale 
   $\mu/\kappa$. }
\end{minipage} \ \hfill
\begin{minipage}[t]{68mm}
  \hspace{-2em}
  \resizebox{1.19\textwidth}{!}{\includegraphics{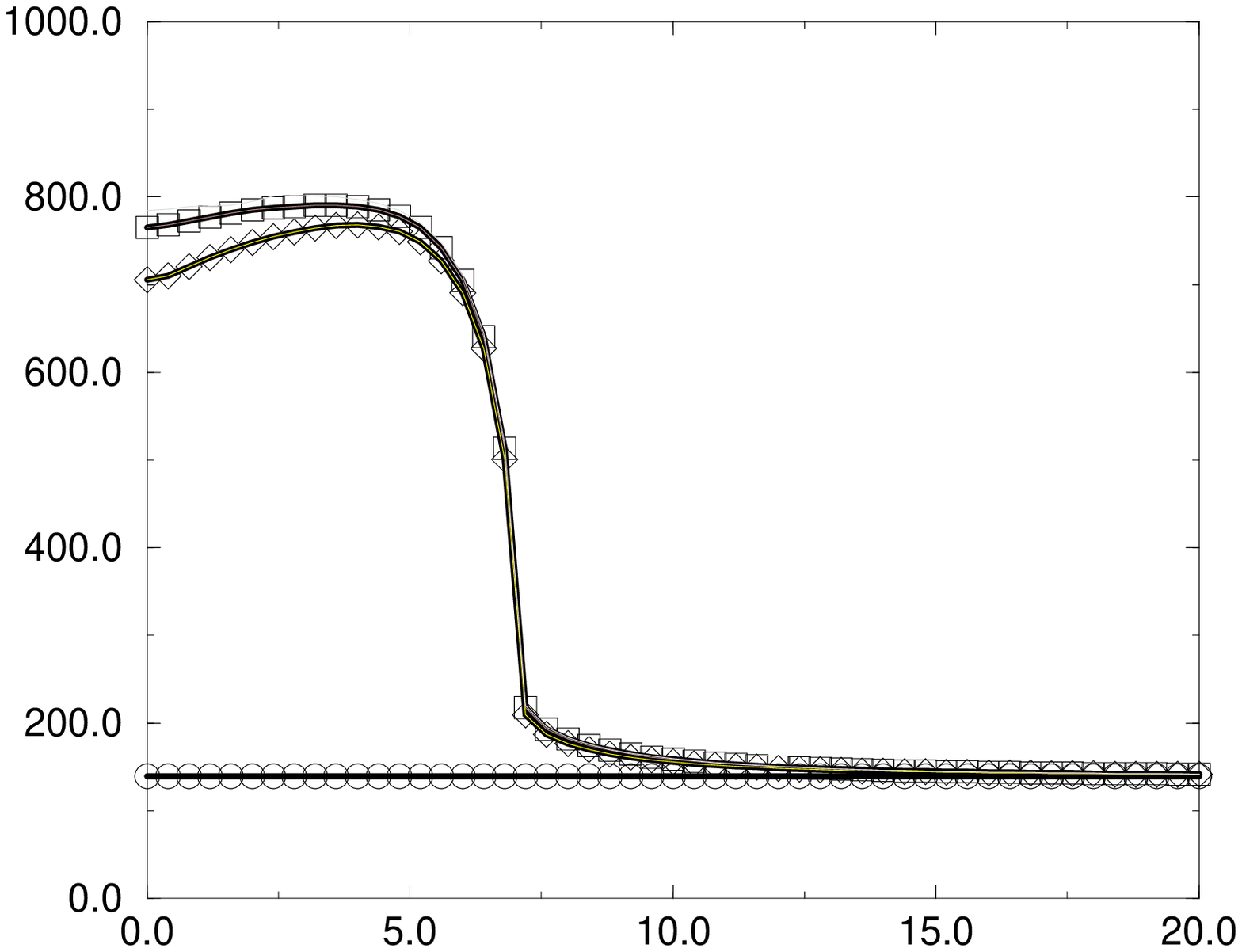}}
\caption{\label{fig:5.2}  
   Spectrum of $\pi^+$: The first three singlet-s states in MeV 
   versus $\mu/\kappa$.}
\end{minipage} 
\end{figure}

For carrying out this programme in practice, 
I need an efficient tool for solving Eq.(\ref{eq:20m}). 
Such one has been developed recently \cite{BieIhmPau99}.
I outline in short the procedure for the special case
$m_1=m_2=m$.
I change integration variables from $x$ to $k_z$ 
by the Sawicki transformation and substitute
\begin{equation}
   x(k_z) = \frac{1}{2} +
   \frac{k_z}{2\sqrt{m^2 + \vec k_{\!\perp}^{\,2} + k_z^2}}
,\hskip3em
   \psi(x,\vec k _{\!\perp}) =
   \ \frac{\phi(k_z,\vec k _{\!\perp})}{\sqrt{x(1-x)}} 
.\label{peq:6}\end{equation}
The variables $(k_z,\vec k_{\!\perp})$ are 
collected in a 3-vector $\vec p$,
with all further details in the Compendium \cite{Com00}.
Eq.(\ref{eq:20m}) becomes then
\begin{eqnarray}
   M^2\phi(\vec p) &=& \left[4m^2+4\vec p^{\,2}\right]
   \phi(\vec p)
\\ &-& 
   \frac{4\alpha}{3\pi^2}\int\!\! d^3\vec p\,' 
   \left(\frac{2m}{(\vec p - \vec p\,')^{\,2}} + \frac{1}{m}
   \frac{\mu^2}{\mu^2+(\vec p - \vec p\,')^{2}} \right)
   \phi(\vec p\,')
.\nonumber\end{eqnarray}
For the present purpose it suffices to restrict to 
spherically symmetric \(s-\)states
and to apply Gaussian quadratures with 16 points.
On an alpha work station it takes a couple of micro-seconds 
to solve for the spectrum of a particular case.
Fixing the up and the down mass to \( m_u = m_d=m \)
(the codes work with dimensionless units)
\begin{equation}
   m=1.16\ \kappa=406\ \mathrm{MeV},  \hskip3em 
   \kappa= 350\ \mathrm{MeV} 
,\label{peq:7}\end{equation}
the calculated pion mass can be made to agree within eight digits
with the experimental value \cite{PDG98}.
For every value of $\mu$ one obtains thus an $\alpha_\mu$ which
is displayed in Fig.~\ref{fig:5.1}.
For every value of $\mu$ (thus $\alpha_\mu$),
I get a whole spectrum, as plotted in Fig.~\ref{fig:5.2}. 
The lowest state corresponds to the the $\pi^+$ 
and stays nailed fixed to the empirical value.
Much to my surprise, the second and the third 
(as well as the higher ones) expose an extremum. 
At the extremum, I fulfill Strutinsky's requirement
\begin{equation}
  \left.\frac{d}{d\mu} M_n^2(\mu)\right\vert_{\mu=\mu_0} = 0
  ,\hskip2em\mathrm{with}\quad
   \mu_0\simeq 3.8\ \kappa,\quad\alpha_{\mu_0}= 0.6904
.\label{peq:8}\end{equation}
The regularization scale $\mu$ determines itself from the solution!

Here comes the next problem: How should I determine the quark mass?
In Eq.(\ref{peq:7}), I selected a particular value of $m$; 
I could have taken
any other and would have gotten $\mu(m)$ and $\alpha_{\mu}(m)$.
I obviously need a second empirical datum. 
Which one?
I could choose the first excited state of $\pi^+$;
but that is not known with sufficient accuracy for my purposes 
and, moreover, there are uncertainties of interpretation, 
see Morningstar \cite{Mor00} and Llanes-Estrada \cite{Lla00}.
I could take the mass of the $\rho^+$;
but the present $\uparrow\downarrow$-model is not 
particularly suited for vector mesons.
I remain thus with the pions size.
Knowing the wavefunction, see below, 
I could calculate the form factor \cite{BroPauPin98} 
and thus the exact root-mean-square radius.
But I replace this non-trivial calculation by the
the following very drastic but simpler construction:
I parametrize the numerical results by a fit function
with one free parameter ($p_a$) to be,
\begin{equation}
  \Phi_a(p) = \frac{10}{\left(1+\displaystyle\frac{p^2}{p_a^2}\right)^2}
  ,\hskip3em\mathrm{with}\quad
  p_a=1.471\ \kappa
.\label{peq:12}\end{equation}
\begin{figure} [t]  
\begin{minipage}[t]{65mm}
  \hspace{-2em}
  \resizebox{1.2\textwidth}{!}{\includegraphics{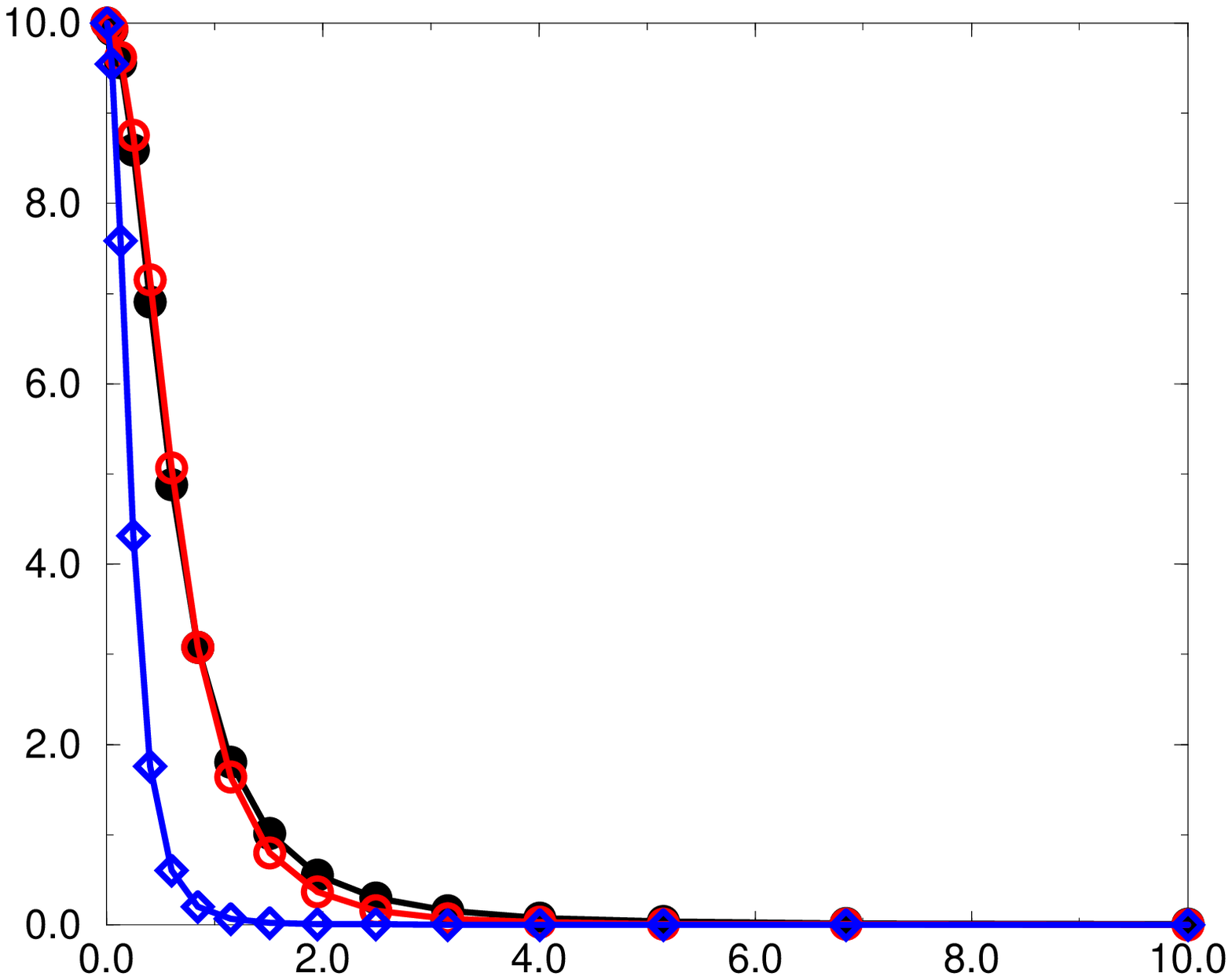}}
  \caption{\label{fig:6.1}  
   The pion wave function $\Phi(p)$ 
   is plotted versus $p/(1.552\kappa)$
   in an arbitrary normalization. 
   The filled circles indicate the numerical results,
   the open circles the fit function.
   The diamonds denote the pure Coulomb solution.}
\end{minipage} \ \hfill
\begin{minipage}[t]{65mm}
  \hspace{-2em}
  \resizebox{1.2\textwidth}{!}{\includegraphics{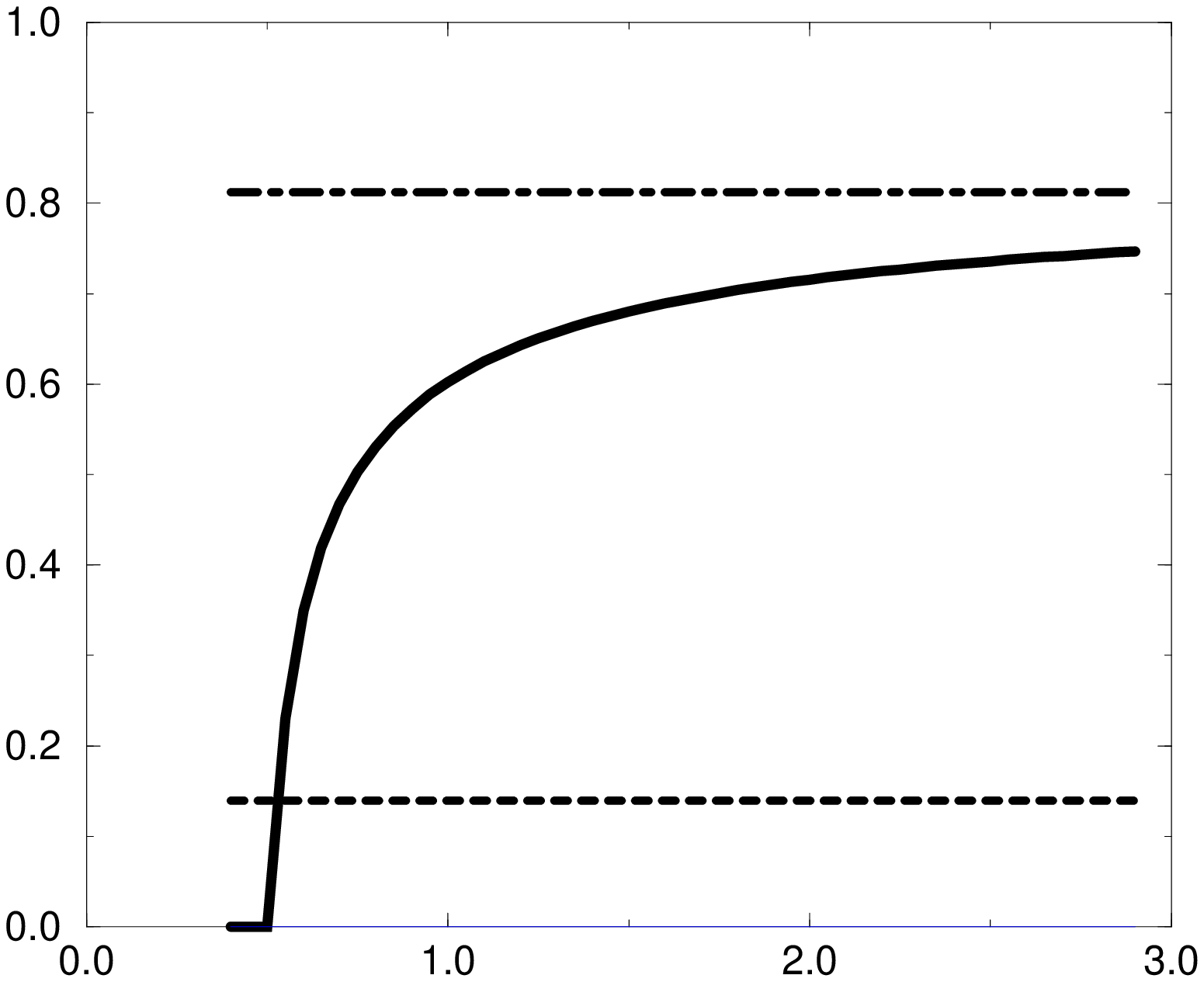}}
  \caption{\label{fig:6.2} 
   The relativistic potential energy $W(r)$ for the pion is plotted 
   in GeV versus the radius in fm (solid line).
   The pion mass is indicated by the dashed, 
   and the ionization threshold by the dashed-dotted line.
   See Eq.(\protect{\ref{peq:16}}).}
\end{minipage} 
\end{figure}
The quality of this fit can be judged by Fig.~\ref{fig:6.1}.
I Fourier-transform this function to configuration space, giving 
$\psi(\vec r) \sim e^{-p_a r}$, and calculate trivially 
\begin{equation}
    <r^2>^\frac{1}{2}_{\phi} = \frac{\sqrt{3}} {p_a} = 0.663 \mathrm{fm}
.\label{peq:14}\end{equation}
Finally, I vary $m$ until I get about agreement with
$<r^2>^\frac{1}{2}=0.657$ fm, 
the experimental value \cite{PDG98}. 
This exhausts all freedom in choosing the parameters of Eq.(\ref{eq:20m}).
Of course, the drastic assumption behind Eq.(\ref{peq:14}) 
has to replaced in the future by the correct 
prescription \cite{MukPau00}.
Actually, I had started the procedure with $m=1.00\ \kappa$.
Eq.(\ref{peq:7}) quotes the result after iteration.
After inserting the inverse Sawicki transformation 
Eq.(\ref{peq:6}) into the fit Eq.(\ref{peq:12}), 
I obtain 
\begin{eqnarray}
   \psi(x,\vec k _{\!\perp}) =
   \frac{1}{\sqrt{x(1-x)}} 
   \frac{\mathcal{N}}{\left(1+\displaystyle
   \frac{m^2\left(2x-1\right)^2+\vec k_{\!\perp}^{\,2}}
   {4x(1-x)\ p_a^2}
   \right)^2} 
.\label{peq:15}\end{eqnarray}
This completes my goal: I have a pion with the correct
mass and size, and I have an analytic expression for its
light-cone wave function.

For the first time in my life I see in Eq.(\ref{peq:15}) 
a genuine light-cone wave function, 
which is directly related to the QCD-Lagrangian
(although only after quite a few approximations and 
simplifications).
It could be used thus as a baseline for calculating the higher
Fock-space amplitudes, as explained in \cite{Pau99b}.
It differs from the literature \cite{leb80} by the preceding 
factor. 
I like to emphasize that regulated delta-interaction,
the Yukawa potential in Eq.(\ref{eq:20m})
at the scale $\mu_0$, acts like a genuine Dirac-delta function:
it pulls down essentially only one state, the pion,
in analogy with the pairing model \cite{BraDamJen72}. 
The others are left more or less unperturbed at their Bohr values,
see Fig.~\ref{fig:5.2}.
I like to emphasize further that I am not in conflict 
with the usual perturbative inclusion of the hyperfine interaction.
For sufficiently large values of $\mu$, say for $\mu \gg 10$,
see Fig.~\ref{fig:5.2}, the coupling constant decreases strongly
and the spectrum becomes more and more the familiar Bohr spectrum 
with a small hyperfine shift, such that 
it can be calculated perturbatively, indeed.
In a future and more rigorous solution of the full equation
I expect that the excited states can be disentangled into
almost degenerate singlets and triplets, which in turn can be
interpreted as an excitation of the pseudo-scalar pion,
or the ground state of a vector meson, respectively.
In any case, the first excited state of the simple model correlates
very well with the mass of the $\rho^+$ and the other vector mesons,
see below.

\section{Extensions and conclusions}

I can apply the same procedure also to the other mesons.
Once I have the up and down mass, I determine
the strange, charm and bottom quark mass 
by reproducing the masses of the 
$K,^-$ $D^0$ and $B,^-$ respectively. This gives
$ m_s = 508$, $ m_c =1666$ and $ m_b =5054\mbox{ MeV}$.
All these numbers as well as those in Tables~\ref{tab:2.2}
and \ref{tab:1.2} are rounded for convenience; 
the calculations or the experiments 
are available with greater precision.
This fixes all parameters which there possibly are.
I generate with them all off-diagonal pseudo-scalar mesons 
and compile their mass  in Table~\ref{tab:2.2}. 
The corresponding wave functions are also available but not shown.
In view of the simplicity of the model,
the agreement with the empirical values 
in Table~\ref{tab:1.2} is remarkable. 
The mass of the first excited states in Table~\ref{tab:2.2} 
correlates astoundingly well with the associated vector meson
in Table~\ref{tab:1.2}.

\begin{table} [t]
\begin{minipage}[t]{68mm}
\caption{\label{tab:2.2}  
   The calculated mass eigenvalues in MeV. 
   Those for singlet-1s states are given in the lower,
   those for singlet-2s states in the upper triangle.}
\begin{tabular}{c|rrrrrr} 
     & $\overline u$ & $\overline d$ 
     & $\overline s$ & $\overline c$ & $\overline b$ \\ \hline
    $u$ &         &      768&      871&     2030&     5418 \\
    $d$ &      140&         &      871&     2030&     5418 \\
    $s$ &      494&      494&         &     2124&     5510 \\
    $c$ &     1865&     1865&     1929&         &     6580 \\
    $b$ &     5279&     5279&     5338&     6114&          
\end{tabular}
\end{minipage} \hfill
\begin{minipage}[t]{68mm}
\caption{\label{tab:1.2}  
   The empirical masses of the flavor-off-diagonal physical mesons in MeV.
   The vector mesons are given in the upper, the scalar mesons
   in the lower triangle.}
\begin{tabular}{c|rrrrrr} 
     & $\overline u$ & $\overline d$ 
     & $\overline s$ & $\overline c$ & $\overline b$ \\ \hline
 $u$ &      & 768  & 892  & 2007 & 5325 \\ 
 $d$ & 140  &      & 896  & 2010 & 5325 \\ 
 $s$ & 494  & 498  &      & 2110 &  --- \\ 
 $c$ & 1865 & 1869 & 1969 &      &  --- \\ 
 $b$ & 5278 & 5279 & 5375 &  --- &      \\ 
\end{tabular}
\end{minipage} 
\end{table}

After so many years, I am forced to 
conclude that the pion looks 
quite a bit different than told in the literature.
Although the above picture is kind of wood-carved,
the previous work with Kalloniatis and Pinsky lets appear it unlikely
that the zero modes, once included, change the picture drastically.
I have found no evidence that the vacuum condensates are important.
I conclude that the pion is describable by a QCD:
The very large coupling constant in conjunction 
with a very strong hyperfine (spin-spin-) interaction
makes it a ultra strongly bounded system of constituent quarks.
More then 80\% 
of the constituent quark mass is eaten up by binding effects.
No other physical system has such a property.
But this pion is not the one I was dreaming of when this work began.

The theory has three parameters.
Two are determined by the pions mass and size,
and one determines itself from the theory.
In the lack of further empirical data, 
I have no independent check for the above strong statements,
except that the first excited state
should roughly be degenerate with the mass of the $\rho$-meson.
At present I am working on a check \cite{MukPau00} whether 
Eq.(\ref{peq:15}) is consistent with Ashery's experiment \cite{Ash00}.

The question of confinement is closely related to a the potential,
the strong potential $V_s(r)$, but in the present approach 
the relation is subtle.
One of the great advantages of light-cone quantization is the
additivity of the free part and the interaction.
Because of the dimensions of invariant mass squares the relation
of potential and kinetic energy is somewhat hidden.
What I can do, however is to Fourier transform Eq.(\ref{eq:20m})
and to interpret the operators $\vec p$ and $\vec q$
as the classical analogue of momentum and position.
Instead of an Hamiltonian $H(q,p)$ I then have
an invariant mass-squared $M^2(q,p)$, 
but for the argument that does not matter.
I thus have
\begin{eqnarray}
    M(r,p) &=& m_s\sqrt{1 + \frac{p ^{\,2}}{m_s m_r} + 
    \frac{2}{m_s}V_s(r)}
,\hskip0.3em
    V_s(r) \equiv \frac{4\alpha}{3r} 
    \left(1+\frac{\mu^2}{m_sm_r}\mathrm{e}^{-p_a r}\right)
.\label{peq:16}\end{eqnarray}
The expansion into  rest mass + kinetic energy + potential energy,
\begin{equation}
    M(r,p)\simeq m_s + \frac{p ^{\,2}}{2m_r} + V_s(r)
,\end{equation}
is justified however only for $p ^{\,2}\ll {m_s m_r}$, {\emph and}
$V_s(r)\ll {m_s}$, which can not be satisfied 
for all $r$. 
I therefore propose to introduce a
{\em relativistic potential energy} by $W(r)\equiv M(r,p=0)$
as the closest relativistic analogue 
to a non-relativistic potential $V_s(r)$,
which we all have in the back of our mind.
This function is plotted in Fig.~\ref{fig:6.2}.
It vanishes at $r_0=0.5194$~fm and the classical turning
point is $r_t=0.5300$~fm. 
Since the sum of the quark masses is 812~MeV,
the ionization threshold occurs at 542~MeV: 
In order to liberate the quarks in the pion,
one has to invest more than three times the rest mass of a pion;
I call this \underline{practical confinement}. 

I contrast these results to Lattice Gauge Calculations.
It is not generally known that LGC's have considerable
uncertainty to extrapolate their results down to such
light mesons as a pion; it exceeds their grid sizes.
It is also not generally known that Lattice Gauge Calculations
get \emph{ always strict and linear confinement}
even for QED, where we positively know the ionization threshold.
The `breaking of the string', or in a more physical language,
the ionization threshold is one of the hot topics
at the lattice conferences \cite{Schilling2000}.
Moreover, in order to get the size of the pion,
thus the form factor, requires another generation of 
computers and physicists to run them.

I contrast these results also to phenomenological approaches.
They usually do not address to get the pion,
and for the heavy mesons, where they are so successful,
a good phenomenological model has quite many parameters,
in any case more that the above canonical ones.
A detailed comparison and systematic discussion of the
bulky literature can however be postponed, until
we are ready to solve the full Eq.(\ref{eq:1})
and do not restrict ourselves to the caricature
of the QCD-interaction, the $\uparrow\downarrow$-model. 

I contrast these results, finally, to the 
Nambu\--Jona\--Lasio\--based models which are so successful
in accounting for the isospin.
I do not even think of quoting from the huge body of literature 
but mention in passing that the NJL-models are not renormalizable,
have no relation to QCD, and deal mostly with the very
light mesons. They just about can handle strangeness and
break down for the heavy flavors.

In conclusion I may state that not a single model can describe
quantitatively \emph{ all mesons} from the $\pi$ up to the $\Upsilon$
from a common point of view, from QCD.
By solving the $\uparrow\downarrow$-model I provide 
at least some evidence that Eq.(\ref{eq:1})
can eventually give that, perhaps sometimes in the future.

\textbf{Acknowledgement.}
I thank Susanne Bielefeld for her unselfish help
during the past three years.

 
\end{document}